\documentclass[a4paper,11pt]{article}
\usepackage{float} 
\usepackage{graphicx} 
\usepackage{tabu} 
\usepackage{url} 
\usepackage{hyperref} 
\usepackage[table,xcdraw]{xcolor}
\usepackage[toc,page]{appendix}
\usepackage{bbding} 
\usepackage{soul}
\usepackage{seqsplit}
\usepackage{threeparttable} 
\usepackage{float} 
\usepackage[table,xcdraw]{xcolor}

\title{\Large \bf Beyond Cookie Monster Amnesia: Real World Persistent Online Tracking}
\small
\author{
	{\rm Nasser Mohammed Al-Fannah}\\
	Information Security Group \\
	Royal Holloway, University of London \\
	\href{mailto:nasser@alfannah.com}{nasser@alfannah.com}
	\and
	{\rm Wanpeng Li}\\
	School of Computing, Mathematics and Digital Technology \\	Manchester Metropolitan University \\
	\href{mailto:W.Li@mmu.ac.uk}{W.Li@mmu.ac.uk}
	\and
	{\rm Chris J Mitchell}\\
	Information Security Group \\
	Royal Holloway, University of London \\
	\href{mailto: me@chrismitchell.net}{me@chrismitchell.net}
} 

\date{}
\begin{document}
	\maketitle
	\begin{abstract}
		\textit{Browser fingerprinting} is a relatively new method of uniquely
identifying browsers that can be used to track web users.  In some ways it is
more privacy-threatening than tracking via cookies, as users have no direct
control over it.  A number of authors have considered the wide variety of
techniques that can be used to fingerprint browsers; however, relatively little
information is available on how widespread browser fingerprinting is, and what
information is collected to create these fingerprints in the real world.  To
help address this gap, we crawled the 10,000 most popular websites; this gave
insights into the number of websites that are using the technique, which
websites are collecting fingerprinting information, and exactly what
information is being retrieved.  We found that approximately 69\% of websites
are, potentially, involved in first-party or third-party browser
fingerprinting.  We further found that third-party browser fingerprinting,
which is potentially more privacy-damaging, appears to be predominant in
practice.  We also describe \textit{FingerprintAlert}, a freely available
browser extension we developed that detects and, optionally, blocks fingerprinting attempts by visited websites.
	\end{abstract}
	
	\section{Introduction}
	A number of authors have discussed the very wide variety of readily
available attributes collectable by websites from a visiting browser, enabling
websites to uniquely identify browsers and potentially track them; this is
known as \textit{browser fingerprinting} \cite{forget,unique,3rd,monster}.
Although the range of retrievable attributes, as well as methods for retrieving
them, have been widely discussed, relatively little has been published
regarding the real-world prevalence of browser fingerprinting, who is deploying
it, and the types of attributes collected to achieve it.  This issue clearly
merits further investigation, and has motivated the work described.

Browser fingerprinting is becoming an increasingly serious privacy concern
despite some apparently benign applications (see Section \ref{bf}).  Its
virtually permanent nature\footnote{Some browser attributes change over time
(e.g.\ browser version) but uniquely identifying browsers is usually still
possible \cite{stalker}, and uniquely identifying the hosting platform is also
possible if a different browser is used \cite{cross}.} is something that might
be subject to future regulation, much as the use of cookies has recently
received the attention of regulators in Europe.  Its use is virtually invisible
to users and there is no direct way of preventing it.  Moreover, we found that
the four browsers used by more than 88\% of web users\footnote{The most
commonly used browser data was retrieved from \url{
https://www.netmarketshare.com/browser-market-share.aspx} [accessed on
01/07/2018].} (i.e.\ Chrome, Internet Explorer, Firefox and Edge) do almost
nothing to help mitigate fingerprinting\footnote{Firefox has a limited set of
options to thwart fingerprinting.}, alert the user to its occurrence, or even
provide information about it in user help documents. 	

We examined the fingerprinting behaviour of the 10,000 most visited websites.
We aimed to discover how many websites deploy browser fingerprinting, whether
directly or through third-parties.  We also examined which attributes are
collected. Further, to help raise awareness of this issue, we developed a
browser extension that alerts users whenever a visited website attempts to
fingerprint their browser; users can also opt to enable a fingerprinting
blocking feature.

The remainder of the paper is organized as follows.  Section~\ref{background}
describes tracking and browser fingerprinting, and reviews relevant prior art.
In Section~\ref{method} the collection of data from 10,000 websites is
described; the results obtained are reported in Section~\ref{results} and
analysed in Section~\ref{analysis}. In Section~\ref{relation} we discuss the
relationship with the prior art. Section~\ref{addon} describes the
\textit{FingerprintAlert} extension, and the paper ends with discussion and
conclusions in Section~\ref{conclusions}.

\section{Background} \label{background}

\subsection{Online tracking}

Online tracking (or web tracking) is the process of monitoring a user's online
activities; entities that perform tracking are known as \textit{trackers}
\cite{raider}.  The methodology used in our study, like that of many other
studies, cannot conclusively determine if a website is actually tracking users;
we simply observe whether they collect attributes from browsers that would
allow them to track via browser fingerprinting. In line with common usage, we
refer to recipients of fingerprintable data (whether first- or third-party) as
trackers.

In practice, the most common motive for online tracking is to enable online
behavioural advertising.  This describes the practice by web advertising
companies of tracking users' online activities in order to display personalised
and targeted advertisements \cite{creepy}.  Additionally, tracking is used as a
tool for market research \cite{3rd}. There are two main approaches to online
tracking --- stateful tracking involving the use of cookies\footnote{A web
cookie is a small amount of data sent by a website as part of an HTTP response
and then stored by the browser. The browser then provides the contents of the
cookie back to the same server in subsequent HTTP requests \cite{cookie}.}, and
stateless tracking, including the use of browser fingerprinting \cite{3rd} as
defined in Section~\ref{bf}.  In this paper, following the seminal work
of Eckersley \cite{unique}, we focus on the latter. 	

In some ways, browser fingerprinting is a more reliable method of tracking than
the use of cookies \cite{hidden}, and it appears that browser fingerprinting is
increasingly being used for this purpose.  Unlike browser fingerprinting,
cookies are stored on user devices and so can be controlled or deleted by
users. In particular, the use of a \textit{private browsing
mode}\footnote{Modes of this type, which have various names, are intended to
enhance the privacy properties of the browser \cite{private}.} as provided by
many browsers, whilst limiting the use of cookies does very little to protect
users against browser fingerprinting \cite{equal}.  Furthermore, while modern
browsers provide a user-selectable \textit{Do Not Track} option, this
apparently does not prevent widespread tracking \cite{detective}. 	

\subsection{Browser Fingerprinting} \label{bf}

Browser fingerprinting enables user web activity to be tracked. It relies on
learning properties of a browser and its host platform, including both hardware
properties and software state (cf.\ the term \textit{device fingerprinting}
\cite{fingerprint}). Browser fingerprinting typically involves a web server
performing some combination of: (a) collecting and analysing information
contained in HTTP request headers, and (b) downloading JavaScript to the
browser which collects and sends back information gathered from browser APIs.
Examples of collected information include: screen resolution, CPU/GPU model,
and names of installed fonts\footnote{A demonstration of the wide range of
information collectable from any browser is available at
\url{https://fingerprintable.org/test}.}.  As in these examples, collectable
attributes relate to both browser and host platform.

Tracking web users has long been possible by using cookies.  However, the
absence of a cookie (e.g.\ because it has been deleted by the user) means that
the device can no longer be tracked \cite{unique}.  By contrast, browser
fingerprinting requires no files to be stored on the user's device, its
effectiveness partly depends on the browser, and users have virtually no control over it
\cite{equal}. It can be used for tracking web users by creating a unique ID
derived by combining collected attributes \cite{beast}.

Four widely discussed uses of browser fingerprinting are: targeted advertising
\cite{detective,raider}; social media sharing \cite{raider,privacy}; analytics
services \cite{detective,raider}; and web security \cite{detective,shpf}. Of
course, browser fingerprinting has other uses, e.g.\ to act as a second layer
of authentication \cite{unique} or to enhance the effectiveness of CAPTCHAs
\cite{harder}. However, even in these cases the server gets the benefit, and
the user is often not informed that fingerprinting is in use \cite{analysis}.
Determining the exact reason(s) why a website deploys browser fingerprinting is
extremely difficult.

Browser fingerprinting websites perform it either as a first-party or a
third-party (or both).  That is, a website may download JavaScript to the
browser, which can send the collected attributes back to either its own site
(first-party fingerprinting) or to a third-party site (third-party
fingerprinting) \cite{defend}.  It is even possible that some website operators
are not aware that a third-party is performing browser fingerprinting via their
website \cite{million}.  This could arise because third-party fingerprinting
sites typically provide client websites with the JavaScript which collects and
sends the attributes used for fingerprinting and in return, the third-party
site provides a range of services to the client website (e.g.\ data analytics
or social plugins).  As a result, some website operators may not know what data
the third-party JavaScript collects from user browsers, or what it might be
used for.

In the context of tracking, first-party fingerprinting gives relatively little
information to a website --- it merely enables multiple visits by the same
browser to be linked, and gives no information about other visited websites.
If the user identity is known by other means (e.g.\ because the user logs in)
it can also indicate when this user is employing multiple devices
\cite{detective}.  Third-party fingerprinting, on the other hand, is much more
privacy-damaging in that it enables browsers (and hence users) to be tracked
across multiple websites.  Later in this paper we report on the websites that
perform the majority of third-party tracking. 	

\subsection{Previous Work} \label{previous}

Back in 2010, Eckersley \cite{unique} first described how the collection of a
range of apparently trivial and readily-available browser attributes, such as
time zone, screen resolution, set of installed plugins, and operating system version, could be combined to uniquely identify a browser; he
gave this process the name browser fingerprinting.  Since then, many other
authors, including Mowery et al.\ \cite{javascript,canvas}, Boda et al.\
\cite{via}, Olejnik et al.\ \cite{battery}, Fifield et al.\ \cite{metric},
Takei et al.\ \cite{css} and Mulazzani et al.\ \cite{engine}, have described a
range of ways of enhancing its effectiveness.  In parallel, and motivated by
the threat to user privacy posed by browser fingerprinting, a number of
authors, e.g.\ Nikiforakis et al.\ \cite{lies}, Fiore et al.\ \cite{fake} and FaizKhademi et al.\ \cite{guard} have proposed ways of limiting its effectiveness. 	

The BrowserLeaks website (\url{https://www.browserleaks.com}) and Alaca et al.\
\cite{augment} catalogue a wide range of types of information that could be
used for browser fingerprinting.   Pathilake et al.\ \cite{classification} have
also classified some of the most widely used methods for fingerprinting.
Browser fingerprinting is clearly very effective; for example, in a large-scale
study, Laperdrix et al.\ \cite{beast} observed that an average of 86\% of
desktop and mobile browsers possess a unique fingerprint; other studies
\cite{unique,javascript} have reported similar results (80--90\%).  It is
important to note that some of the attributes that can be used for
fingerprinting vary between desktop and mobile platforms; as a result the
efficiency of fingerprinting also varies between platform types \cite{beast}.
For example, a device model name can be retrieved from a mobile browser
\textit{user agent} but not from its desktop counterpart. 	

We conclude this brief review of the prior art by summarising previous work
with a similar scope to that of this paper, namely examining the prevalence and
nature of browser fingerprinting.  In 2015, Libert \cite{hidden} published the
results of a study of third-party HTTP requests utilized for browser
fingerprinting.  Acar et al.\ \cite{detective} performed a large-scale study of
fingerprinting focussing mainly on detection by whether a site examined the set
of installed fonts. More recently, Le et al.\ \cite{accurate} followed a
similar approach, but based detection on use of the canvas API rather than the
installed fonts. Englehard et al.\ \cite{million} performed one of the most
comprehensive studies in this area, although they focussed on tracking in
general and not just on stateless (fingerprinting-based) tracking. Englehardt
et al.\ examined the JavaScript downloaded by websites to browsers, a
potentially rich source of information, using their own tool, OpenWPM.
According to the authors, this tool performs better than many other similar
tools such as FPDetective \cite{detective}. However, the use of automated tools
to examine JavaScript has limitations, in that tools can only look for scripts
they are programmed to identify, regardless of the nature of data collected by
a tracker.  Metwalley et al.\ \cite{horde} also examined the prevalence of
tracking; however, they looked at a relatively limited number of websites (500)
and aimed to detect all types of online tracking via passive measurements
rather than looking specifically at fingerprinting. 			

\subsection{Motivation}

Despite the fact that browser fingerprinting has been extensively studied,
relatively little information has appeared on its prevalence and the browser
attributes that are collected in practice.  To the authors' knowledge, no other
study has listed all the browser fingerprinting attributes that are collected
by a large set of real-world websites.  This observation motivates the work
described in the sequel, in which we describe a study of the fingerprinting
behaviour of the 10,000 most popular websites.  Unlike the work of Englehardt
et al.\ \cite{million} and Acar et al.\ \cite{detective}, we chose not to
examine the JavaScript itself, but instead monitor the data that is actually
transferred back from the browser. While adopting a somewhat similar method,
the scale of the study is more than an order of magnitude larger than the study
of Metwalley et al.\ \cite{horde}. 	

One important motive for understanding better the prevalence and nature of
browser fingerprinting is to help in developing tools that inform the user about
fingerprinting, and also enable users to exert control over the degree to which
fingerprinting is possible.  To this latter end, in Section \ref{addon} we
describe \textit{FingerprintAlert}, a browser extension developed as part of the
study, which makes users aware whenever a website is collecting information
usable for browser fingerprinting.  It also allows all detected fingerprinting
to be blocked. 	

\section{Data Collection Methodology} \label{method}

\subsection{Data Gathering}

The main objectives of the data collection exercise were to assess the number
of websites performing browser fingerprinting, and what types of data are being
collected for this purpose.  To achieve our objectives, we decided to crawl a
large number of well-used websites and to test their data gathering behaviour.  We chose 10,000 sites, as this seemed both sufficiently many to generate
representative results, and also a manageable number so we could analyse the
considerable volumes of data generated.  We only looked
at the data transmitted, rather than analysing the downloaded JavaScript, for
two main reasons: manual analysis of JavaScript on this scale was infeasible,
and automated analysis, as noted above, has limitations. Moreover, the data
that is sent was the key issue of concern for us, not so much how it is
gathered. 	

We used a simple method to decide whether a web server is performing browser
fingerprinting.  To try to ``normalize" web server behaviour, we looked only at
the interactions that occur when a browser initially visits the homepage of the
website, rather than other information gathering exercises that might occur
(e.g.\ when a user tries to log in). So, a website that sends any
fingerprinting browser attributes back to its, or a third-party, server at a
first visit has been \textit{deemed} to be engaged in browser fingerprinting;
the precise criterion used to decide whether a site is fingerprinting is given
in Section \ref{processing}. 	
	
\subsection{Experimental Set Up} \label{experiment}

In order to select which websites to crawl, we retrieved the top 10,000
websites from the freely available Majestic list of the one million most
visited websites\footnote{Majestic is a website specializing in web usage
statistics, and provides a daily-updated list of the top one million websites,
\url{https://majestic.com/reports/majestic-million}  [accessed on
09/10/2017].}.  We wrote a program to crawl the homepages of these websites to
discover if they employ browser fingerprinting techniques at the point when the
website is first loaded (i.e.\ prior to any interaction).  This of course means
that we missed websites that employ interaction-triggered fingerprinting.  The
crawler was created using Selenium WebDriver\footnote{Selenium is open-source
software used to automate browsers for testing purposes --- see \url{
https://www.seleniumhq.org}.}, a Python script, the \textit{FingerprintAlert}
extension, and the Chrome browser (details of the crawler software components and
the device used can found in Appendices \ref{crawl} and \ref{config}).  The
Python script instructs Selenium to visit the 10,000 websites in the list, wait
for each to fully load, and then wait for a further short period before moving
to the next website. 	

The delay is included because, in preparatory work, we manually visited 50
websites on the list and found that some only relayed information after a delay
ranging from one second to several minutes following the full loading of the
page.  Such waits seem likely to be both to allow the various elements of the
web page to be loaded and executed and to take account of dynamic content being
continuously loaded (e.g.\ advertisements). We set the short delay to 3
seconds; this was a fairly arbitrary choice, although it was long enough to
cause a number of websites to transmit data, although not sufficiently long to
make the crawling process significantly more time consuming. 	

The extension collects and stores all data that is relayed from the browser to one
or more web servers using the GET, POST or HEAD HTTP methods\footnote{The
quantity of data that can be relayed using GET or HEAD is very limited, whereas
POST allows the transmission of very large volumes (megabytes) of data.}
\cite{http}, i.e.\ the commonly used means by which information, including
attributes used for fingerprinting, is relayed from browser to server.  Whether
or not the data was sent SSL/TLS-protected, i.e.\ using HTTPS \cite{https}, was
also recorded.


The crawling process took approximately 300 hours to complete.  It took this
long for several reasons, including that some websites took several minutes to
fully load, and that Selenium occasionally crashed.  In such cases, the crawler
was restarted manually, where we re-crawled websites after a
crash to ensure we did not miss any data.

\subsection{Data Processing} \label{processing}

Prior to the full crawling process we initially crawled a smaller sample
(approximately 1,000 of the websites) to test the crawler.  In this process we
indiscriminately collected all data sent (if any) from the browser to web
servers.  Manual examination of the collected data revealed it included
information unrelated to the visiting device or the browser (e.g.\ the URLs of
displayed advertisements), i.e.\ of no interest to this study.  Most
importantly for our purposes, we were able identify fingerprinting attributes
that had unique formats or values (e.g.\ screen resolution: 1920x1080) that
made automatic detection possible. Using these preliminary findings, we
programmed our crawler to automatically detect a set of 17 attributes (as
listed in Appendix \ref{prepopulate}).  The crawler used regular expressions to
examine relayed data and match them with the prepopulated attributes.

The presence of one or more of these attributes in data returned by the browser
was used to determine whether or not a website was engaged in fingerprinting.
This set of 17 attribute types includes many of the attributes whose use for
fingerprinting is most widely discussed, so we believe that the presence or
absence of an attribute of one of these types is a reasonable indicator of
whether fingerprinting is being performed.

However, other attributes are much more complex, and hence are difficult to
automatically identify.  In subsequent manual analysis of the recorded data, we
were able to identify many additional attributes because they were labelled by
name in the captured data.  To perform this task automatically would have been
extremely difficult because some sections of the recorded data were not parsed,
and the substrings of the data that were parsed varied in format
(unsurprisingly given the absence of any standards for data formats for
transferred attributes).

In order to manually identify fingerprinting attributes in the collected data,
we first used publicly available scripts to retrieve a large set of
fingerprinting attributes from the browser that was used to run the experiments
(the scripts we used can be found at \url{https://github.com/fingerprintable}).
We then attempted to match these values with the values in the collected data.
Once we completed the matching, we manually inspected the matches found; this
was necessary to ensure that the matches found were genuine and not
coincidental similarities in strings or numbers.  In most cases the match was
confirmed by finding labels followed by the expected values in the collected
data.

\subsection{Challenges Addressed}

We faced a number of challenges in both implementing crawling and processing
the collected data.  First, websites are unlikely to admit use of browser
fingerprinting, and so we can only attempt to judge their behaviour based on
the types of information retrieved from the browser, and when it was collected.
As mentioned earlier, there is a wide range of attributes that, when put
together, can be used to create a unique device fingerprint.  Identifying and
monitoring all such attributes is very challenging, especially since new
attributes seem to arise frequently (given continuously evolving browser
functionality). Moreover, many websites cause the browser to send a series of
data strings back to the server; automatically, or even manually, identifying
what these data represent is highly non-trivial.  It was not always possible to
parse the data sent since there is no standard for such data transmissions;
indeed, some websites may deliberately obfuscate the data they send. It was
therefore impossible to fully interpret all the data. Fortunately, there are
certain attributes that are easily identifiable because of their special format
and range of values, such as screen resolution (e.g.\ 1920x1080), fonts (e.g.\
Arial), or geolocation coordinates (e.g.\ 51.4167, -0.5667).

It is very difficult to determine the minimum number of attributes needed to
produce a unique fingerprint.  Fingerprint uniqueness depends on many factors,
including the range of values of an attribute, how often it changes, and how
different it is between one browser/platform and another.  As a result, we made
the simplifying assumption that a website is deemed a tracker if it causes a
browser to send at least one of the 17 attributes given in Appendix
\ref{prepopulate}.

As our crawler was Selenium-based, it suffered from the known crashing problem
\cite{million} on certain websites, e.g.\ when it was unable to fully load all
the elements of a website.  In such cases the crawler had to be manually
restarted.  On average, Selenium crashed once in every 155 visited websites.
Moreover, Chrome extensions are limited to 5MB of storage and so, to ensure that
the collected data did not reach that limit, we programmed the crawler to stop
after every 200 visited websites, yielding an average of 3MB of collected data.
However, Selenium usually crashed before reaching the 200-website limit. 		

The 10,000 websites took an average of 19 seconds to fully load.  Our tests
were performed using an Internet connection with a minimum bandwidth of 40
Mbps, and so connection limitations are unlikely to be the reason for the
loading delays.  The time to load a website noticeably increased as we went
through the list of crawled websites, i.e.\ the less popular websites loaded
more slowly.  So, in future similar experiments, we would recommend that
crawlers should not timeout until at least 20 seconds have elapsed. 		

\section{Results} \label{results}

\begin{figure}[htb]
		\centering
		\includegraphics[width=\textwidth]{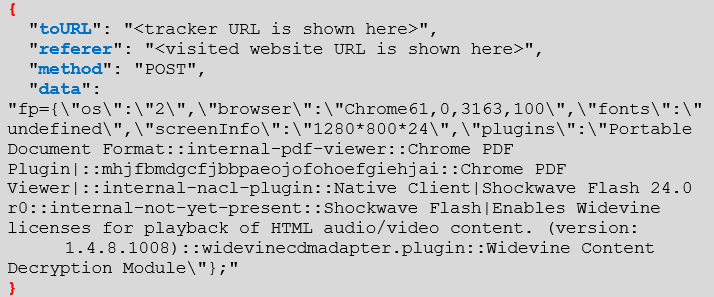}
		\caption{Excerpt of collected data}
		\label{sample}
\end{figure}

The data collected in this study, as well as the tools we used for data
collection and analysis, are available at \url{
https://github.com/fingerprintable}.  The dataset includes the contents of all
HTTP messages sent by and to the crawled websites that attempted
fingerprinting.  This includes the data retrieved from the visiting device
(i.e.\ the device used for data gathering), as well as the domain names of the
sender and receiver of the data.  Figure \ref{sample} shows a sample of a
complete block of data from amongst those collected in our study.

Using a combination of automated parsing and manual inspection, we detected the
transmission of 284 different attribute types.  We further detected 1,914
distinct fingerprinters.  70 websites (i.e.\ 0.7\%) timed out (e.g.\ because
the website did not respond) during the crawling process and thus were fully,
or partially, excluded from our findings.  Overall, 6,876 (68.8\%) of the
crawled websites collected data from visiting browsers (as first- or
third-parties) that could be used for browser fingerprinting.  We refer to such
websites as \textit{fingerprinting websites}; of course, despite the name, the
fingerprinting websites might not actually be using the collected data for
fingerprinting.


Fingerprinting is most commonly performed by third-party sites; 84.5\% of the
6,876 sites collecting data sent it only to third-parties. Of the rest, 2.4\%
were exclusively first-party fingerprinters, with the other 13.1\% using both
first- and third-party data collection.  Over the 6,876 fingerprinting
websites, data was sent to an average of 3.42 domains. The largest number of
different data-collecting websites to which data was sent for a single visited
website was 42.

\begin{figure}[htb]
		\centering
		\includegraphics[width=\textwidth]{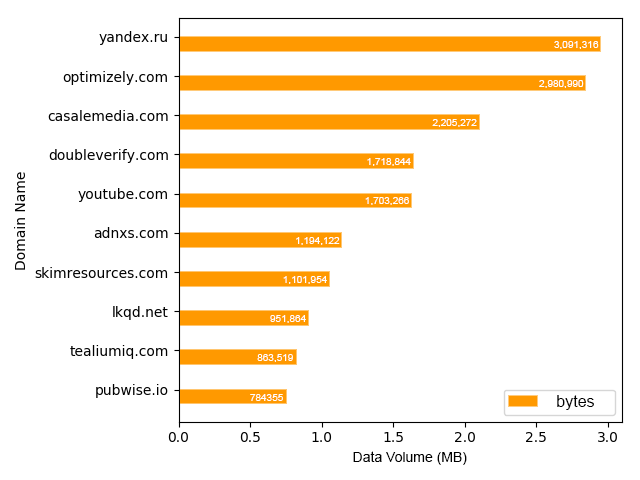}
		\caption{Top 10 fingerprinters in terms of collected data volume per browser}
		\label{size}
	\end{figure}

Fingerprinting websites collected an average of 1.75KB of data.  The
third-party websites that collected the most data were yandex (2.9MB),
optimizely (2.8MB) and casalemedia (2.1MB). Figure \ref{size} shows the top 10
third-party websites in terms of collected data volume for a single visiting
browser. 	

\begin{figure}[htb]
		\centering
		\includegraphics[width=\textwidth]{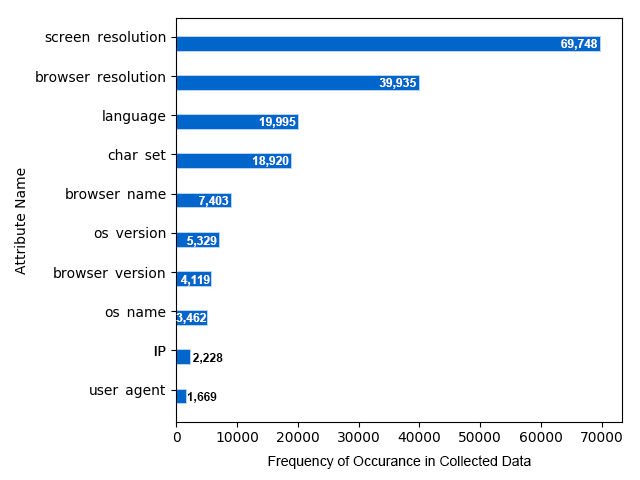}
		\caption{Top 10 collected attributes }
		\label{attributesx}
	\end{figure}

Of the attributes we can automatically detect, the three most frequently
collected were: screen/browser resolution, language, and charset (i.e.\
character encoding). We found that fingerprinters collected, on average, 5 of
the 17 pre-populated attributes.  Figure \ref{attributesx} summarises the 10
most frequently collected attribute types.  The most widely used fingerprinting
third-party was google-analytics\footnote{\url{https://analytics.google.com}}
(see \url{https://github.com/fingerprintable} for a complete list of
fingerprinting third-parties); google-analytics provides web analytics as well
as other web-based services to websites.  DoubleClick\footnote{
\url{https://www.doubleclickbygoogle.com}} (Google's online advertising
service) was the website that collected the largest volume of data overall. 		

As noted above, amongst the collected data we were able to identify 284
fingerprinting attributes, which we divided into six categories (see Table
\ref{attributes}).   The full list of 284 attributes can be found in Appendix \ref{full}. 	

\begin{table*}[htb]
	\small
	\centering
	\caption{Summary of identified fingerprinting attributes}
	\label{attributes}
	\begin{tabular}{|p{6.6em}|p{3.3em}|p{3.7em}|p{2.6em}|p{2.2em}|p{1.4em}|p{3.8em}|p{2.1em}|}
		\hline
		\rowcolor[HTML]{ EFEFEF}
		Attribute Type & \textbf{WebGL}        & \textbf{Features} & \textbf{Media}        & \textbf{Misc.}   & \textbf{IO*} & \textbf{Network}  & \textbf{Total}
		\\ \hline
		Count         & 114          & 66       & 41           & 35        & 20           & 10       & \textbf{286}            	
		\\ \hline
		
	\end{tabular}
	\begin{tablenotes}\footnotesize
		\item *Input/Output
	\end{tablenotes}
\end{table*}

	
\section{Analysis}  \label{analysis}

\subsection{Processing Collected Data}

The crawler logged every website that relayed data if one, or more, of the 17
pre-programmed attributes were detected.  We examined random samples of the
collected data to identify the presence of any false positives.  We found some
HTTP messages that contained data that were incorrectly matched with one of the
17 attributes.  We wrote a script to remove such records (e.g.\ if the string
\verb=1280088.jpeg= matched with the screen resolution width \verb=1280=). This
filtering reduced the number of false positives.  However, in general,
identifying false positives (if any) in the filtered data is non-trivial since
the ability to fingerprint browsers typically depends on both the number
and type of collected attributes.  For example, Mowery et al.\ \cite{canvas}
have demonstrated that the canvas API alone could be enough to
fingerprint a browser, and Laperdrix et al.\ \cite{beast} demonstrated a
seemingly successful method of fingerprinting based on a specific set of just 17 attributes. 	

\subsection{Undetected Fingerprinting}

As noted in Section \ref{experiment}, the crawler only visited the
\textit{homepages} of the 10,000 websites.  Websites we reported as not
deploying browser fingerprinting might nevertheless still be doing so on other
pages.  Moreover, the attribute collection reported here was unprompted (i.e.\
no clicking, cursor movements or typing was involved) except for loading of the
web page.  Through manual visits to selected websites, we found that some
websites only cause the browser to send fingerprinting attributes when there
are further interactions.  Moreover, some websites only retrieved attributes
when a user submits a form or logs in, and such cases would be too complex (if not impossible) to
capture automatically.  The focus of this study is fingerprinting that targets
everyone, including those engaged in casual browsing.

\subsection{Prevalence of Fingerprinting} \label{prevalence}

\begin{figure}[htb]
		\centering
		\includegraphics[width=\textwidth]{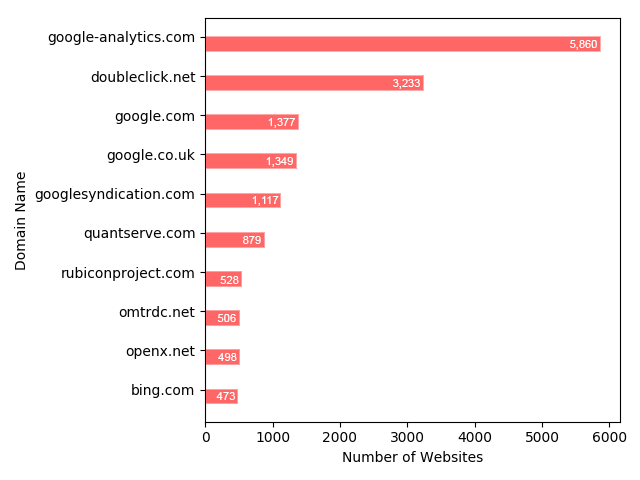}
		\caption{Top third-party fingerprinting domains}
		\label{fingerprinters}
	\end{figure}

Our study confirms the findings of Englehardt and Narayanan \cite{million}
that fingerprinting is commonplace, at least by widely-used websites, and yet
there are a relatively small number of entities actually collecting and
processing attributes (mainly third-party trackers).  Indeed, the top five
third-party fingerprinting domains (see Figure \ref{fingerprinters}) are all
part of a single company, Google Inc.  This finding is consistent with Libert
\cite{hidden}, who found that 78.07\% of the top one million websites send data
to a Google-owned domain.

We found that 68.8\% of the top 10,000 websites are potentially engaged in
fingerprinting, although previous studies yielded rather different results. For
example, in 2013, Nikiforakis et al. \cite{monster} found that only 0.4\% of
the top 10,000 websites deployed fingerprinting. A year later, Acar et al.
\cite{forget} reported that 5\% of the top 100,000 websites deployed browser
fingerprinting using the canvas API\@.  It thus seems likely that both the
prevalence of browser fingerprinting and the number of attributes being
collected for this purpose have significantly increased.

\subsection{Fingerprinting Attributes}

We attempted to find the fingerprinting attributes reported by Alaca et al.\
\cite{augment} and the BrowserLeaks website in the collected data, including
attributes not in the list of 17 attribute types detectable by the crawler.
This gave us an indication of the range of attributes that are collected in the
real world, as opposed to those discussed in the literature, and also helped us
improve the functioning of the extension described in the Section \ref{addon}.

As reported above, we were able to identify the collection of 284 attributes, a
much larger number than those reported by previous studies. This is partly
explained by the fact that previous studies have searched for a smaller number
of attributes; for example Eckersley \cite{unique} and Cao et al. \cite{cross}
looked for just 10 and 53 respectively.  The significantly higher number we
found also seems likely to be a result of the growing use of browser
fingerprinting \cite{detective,monster}, and the fact that we monitored the
HTTP messages transmitted between visited websites and potential trackers as
opposed to detecting the presence of pre-identified fingerprinting scripts, as
previously widely performed. Most of the attributes we were able to identify
are collectable by BrowserLeaks.com. However, BrowserLeaks can also collect
many attributes that we did not find any websites to be collecting, including
many of the browser features collectable by \textit{Modernizr}\footnote{A
JavaScript library that help websites detect the availability of css and html5
features in a visitor's browser \url{https://modernizr.com}}.

\subsection{Deployment of HTTPS}

Some fingerprinting websites do not use HTTPS to send the fingerprinting
attributes which are thus transmitted in plaintext; this is a potentially
significant user privacy threat.  Of the 1,914 distinct fingerprinters we detected, as many as 683 used only HTTP for attribute transmission, 274 mixed
use of HTTP and HTTPS, and the remaining 957 used only HTTPS.  That is, 50\% of
the fingerprinting websites used HTTP at least in some cases for transmitting
what could be construed as personally identifiable information.  Seemingly, the
use of HTTP is more common in less popular websites, as Merzdovnik et al.\
\cite{catch} reported that as many as 60\% of the top 100,000 websites
performing fingerprinting used HTTP.  We identified a fingerprinting website
that used the WebSocket protocol\footnote{It is a relatively new full-duplex
TCP communication protocol \cite{ws}.} as well as HTTP.  These results apply
only to the use of HTTP/HTTPs for transmitting browser attributes, not to
whether or not the visited website uses HTTPS.

\subsection{Fingerprint IDs}

Some websites cause a browser to send a value that is explicitly labelled
\textit{fingerprint} or \textit{fp}, along with fingerprinting attributes.
These values are typically strings of alphanumerics that appear to function as
platform/user identifiers.  Evidently, some first- and third-party trackers
share such user identifiers \cite{across}, allowing them to compile extensive
profiles of users.  This also means that a website or a tracker could acquire
user- or platform-related information without any prior interaction with that
user.  Such ID-sharing practices clearly make browser fingerprinting-based
tracking more privacy-threatening. 	

\section{Relationship to the Prior Art} \label{relation}

Our study, like that of Libert \cite{hidden}, examined HTTP requests; however,
whereas Libert examined only third-party tracking, we also considered
first-party tracking, i.e.\ by the visited website itself.  Moreover, we
focussed on browser fingerprinting and not on tracking via cookies, a topic
that has been extensively examined in the prior art (e.g.\ Felten et al.\
\cite{timing}, Krishnamurthy et al.\ \cite{footprint} and Mayer et al.\
\cite{3rd}).  A further difference between the work described here and several
previous studies, including that of Englehardt et al.\ \cite{million}, is that
they examined the fingerprinting scripts while we examined the data relayed
back to server via HTTP.  Most significantly, and as discussed in Section
\ref{prevalence}, we detected a much higher level of browser fingerprinting
than previously reported; indeed, our results suggest that fingerprinting is
becoming ubiquitous.

Given that this is a rapidly changing and evolving area, it is important to
repeat studies frequently, and so one contribution of our work is to reveal the
current state of the art.  We do not claim that the approach we have adopted is
better than other approaches, but it does have the advantage of being based
purely on the data itself, and not on the many and various scripts that might
be used to fingerprint browsers.  Our study has enabled us to give an up to
date, fairly comprehensive, and large-scale list of the attributes being used
in practice for browser fingerprinting. 	

\section{Browser Extension} \label{addon}

\subsection{Overview}

As part of the research described here, we developed
\textit{FingerprintAlert}\footnote{\url{https://chrome.google.com/webstore/detail/ielakmofegkdlpnlppfikmkbceajdofo}\\
\url{https://addons.mozilla.org/en-US/firefox/addon/fingerprintalert}}, a
browser extension compatible with desktop versions of Chrome and Firefox for both
Windows and macOS.  Based on the preliminary crawling described in Section
\ref{processing}, we programmed the extension to detect the same 17 attributes.
It is activated whenever a web page is loaded, and checks whether any of these
pre-specified attributes are being relayed back to a web server.  If the extension
detects such activity, it displays an alert that includes both the sending and
receiving URLs.  The extension also provides a detailed report of detected
activities, including data relayed and the corresponding destination(s).
Finally, the extension offers a user-selectable option to automatically block
detected fingerprinting attempts.  If selected, an HTTP message including any
of the monitored attributes will be blocked from being relayed back to a remote
server.  Despite only detecting 17 attributes, these attributes are typically
transmitted alongside other attributes which are also blocked, given that they
are in the same HTTP message.

\subsection{Blocking Feature}

Websites typically send collected data in a series of HTTP messages, and
\textit{FingerprintAlert} blocks those messages that contain at least one the
17 attributes.  We found that these attributes are typically transmitted in the
same HTTP message as a large number of other fingerprinting attributes, which
are also blocked as a result.

As with any extension that interferes with browser behaviour, the blocking feature
of \textit{FingerprintAlert} might cause unexpected results or even break some
websites.  To ensure it does not cause significant usability issues, we tested
it on the 50 most visited websites from our list. We enabled the blocking
feature, and spent around two minutes on each website performing actions such
as signing up, logging in and clicking on links.  During the tests we did not
observe any unexpected behaviour or errors except for some glitches on two
websites (e.g.\ unable to load support chat window). Nonetheless,
in the unlikely event that the extension damages a user's experience at a website,
the blocking option or the notifications option can easily be disabled.  The
extension will continue to record detected fingerprinting attempts even if both
these options are disabled.

\subsection{Challenges}

Detecting newer or obscure fingerprinting attributes is an obstacle that faces
all privacy extensions \cite{million}.  Moreover, websites could choose to conceal
transmitted attributes, e.g.\ using encryption, or use fingerprinting
attributes that are not publicly known.  Additionally, it is difficult to
automatically detect all fingerprinting attribute values, as they may be
similar to other data or have no specific set of values.  On the other hand,
detecting and examining scripts executed on websites is likely to be hindered
by changes in code, syntax and execution.  For that reason, the extension notifies
the user if any HTTP message sent to a server is found to contain one or more
of the selected set of 17 attributes.	

\subsection{Other Extensions and Future Improvements}

The extension complements, rather than replaces, other extensions that mitigate
fingerprinting, such as those that monitor and block fingerprinting scripts
(e.g.\ Ghoesrty\footnote{\url{https://www.ghostery.com}} and Privacy
Badger\footnote{\url{https://www.eff.org/privacybadger}}).  The main purpose of
our extension is to make users aware of fingerprinting attempts as they happen and
the identity of domains collecting the fingerprinting data, and as a result
increase their awareness of how widespread such practices are.  The results of
our study could also help in developing new tools designed to thwart
fingerprinting. In the future, we aim to improve \textit{FingerprintAlert} by
increasing the number of automatically-detectable attributes. This can be
achieved by further in-depth examination of the formats and values of
attributes that are currently undetectable. Since the crawler is based on the
extension, any future crawls would also be made more effective by such
improvements.

\section{Discussion and Conclusions}  \label{conclusions}

Cookies are familiar to many users, especially with the introduction of
regulations on their use, such as the so-called cookie
law\footnote{\url{http://ec.europa.eu/ipg/basics/legal/cookies/index_en.htm}
[accessed on 14/04/2018]} covering tracking whether using cookies or any other
technology.  These regulations have caused many websites to announce the use of
cookies.  However, while users can disable local storage of cookies, cookies
can be selectively deleted, and cookies expire, browser fingerprinting is
virtually outside of user control and is much more permanent; it is thus
significantly more threatening to user privacy.

Many authors, e.g.\ Nikiforakis et al.\ \cite{lies} and Torres et al.\
\cite{block}, have described means of reducing the effectiveness of
fingerprinting through browser extensions or by adjusting user-configurable
browser settings.  Previously described extensions typically either hide certain
attributes or fabricate their values.  While such extensions can be helpful, they
also have well-known limitations; exhibiting an unrealistic set of attributes
values is also fingerprintable \cite{monster} and could negatively affect the
browsing experience (e.g.\ if screen resolution values are manipulated).

We have shown that browser fingerprinting is being conducted on a significantly
larger scale than previously reported, involving the transmission of large
volumes of browser and device-specific data to trackers. We also reported on
the large number of fingerprinting attributes collected.  As other authors have
described, browser fingerprinting has significant negative implications for
user privacy, and it is therefore important that the web user community is made
aware of its prevalence and potential effectiveness.  To this end we have
developed \textit{FingerprintAlert}, that informs users when fingerprinting is
occurring and can also block it.  If web user privacy is to be preserved,
fingerprinting technology needs to be made user-controllable so users can limit
the degree to which they are tracked. Our browser extension contributes to this by
providing users with the option to block browser fingerprinting.  In the longer
term it may be necessary for regulators to examine ways of limiting the degree
to which users are tracked using fingerprinting, and/or for browser
manufacturers to find ways of developing browsers that limit how easily one
user can be distinguished from another.

\textbf{Ethical Issues.} Clearly any experiment involving real world websites
raises potential ethical issues.  However, no data relating to individuals were
accessed, no vulnerabilities in websites were discovered or exploited, and all
websites were accessed as intended by their providers.  Websites were crawled
only once, except in cases of a crawler crash where an additional visit was
required.  All the results are publicly available, as described in Section
\ref{results}. 		

	\bibliographystyle{plain}
\bibliography{Cookieless}

\begin{thebibliography}{10}

\bibitem{forget}
Gunes Acar, Christian Eubank, Steven Englehardt, Marc Ju{\'{a}}rez, Arvind
  Narayanan, and Claudia D{\'{\i}}az.
\newblock The web never forgets: Persistent tracking mechanisms in the wild.
\newblock In {\em {ACM} {SIGSAC} 2014, Scottsdale, AZ, {USA}, November 3--7,
  2014.}, pages 674--689. {ACM}, 2014.

\bibitem{detective}
Gunes Acar, Marc Ju{\'{a}}rez, Nick Nikiforakis, Claudia D{\'{\i}}az, Seda~F.
  G{\"{u}}rses, Frank Piessens, and Bart Preneel.
\newblock {FPDetective}: dusting the web for fingerprinters.
\newblock In {\em {ACM} {SIGSAC} {CCS'13}, Berlin, Germany, November 4--8,
  2013.}, pages 1129--1140. {ACM}, 2013.

\bibitem{harder}
N.~M. Al-Fannah.
\newblock Making defeating captchas harder for bots.
\newblock In {\em Computing Conference 2017, London, UK, July 18--20, 2017.},
  pages 775--782. {IEEE} Computer Society, July 2017.

\bibitem{equal}
Nasser~Mohammed Al{-}Fannah and Wanpeng Li.
\newblock Not all browsers are created equal: Comparing web browser
  fingerprintability.
\newblock In {\em Advances in Information and Computer Security --- {IWSEC}
  2017, Hiroshima, Japan, August 30 -- September 1, 2017.}, volume 10418 of
  {\em Lecture Notes in Computer Science}, pages 105--120. Springer, 2017.

\bibitem{augment}
Furkan Alaca and Paul~C. van Oorschot.
\newblock Device fingerprinting for augmenting web authentication:
  classification and analysis of methods.
\newblock In {\em {ACSAC} 2016, Los Angeles, CA, USA, December 5--9, 2016.},
  pages 289--301. {ACM}, 2016.

\bibitem{cookie}
A.~Barth and U.C. Berkeley.
\newblock {{HTTP} State Management Mechanism}.
\newblock {RFC} 6265, {RFC Editor}, April 2011.

\bibitem{via}
K{\'{a}}roly Boda, {\'{A}}d{\'{a}}m~M{\'{a}}t{\'{e}} F{\"{o}}ldes,
  G{\'{a}}bor~Gy{\"{o}}rgy Guly{\'{a}}s, and S{\'{a}}ndor Imre.
\newblock User tracking on the web via cross-browser fingerprinting.
\newblock In {\em {NordSec} 2011, Tallinn, Estonia, October 26--28, 2011.},
  volume 7161 of {\em Lecture Notes in Computer Science}, pages 31--46.
  Springer, 2011.

\bibitem{cross}
Yinzhi Cao, Song Li, and Erik Wijmans.
\newblock (cross-)browser fingerprinting via {OS} and hardware level features.
\newblock In {\em {NDSS} 2017, San Diego, California, USA, February 26 -- March
  1, 2017}. The Internet Society, 2017.

\bibitem{unique}
Peter Eckersley.
\newblock How unique is your web browser?
\newblock In {\em {PETS}2010, Berlin, Germany, July 21--23, 2010.}, volume 6205
  of {\em Lecture Notes in Computer Science}, pages 1--18. Springer, 2010.

\bibitem{million}
Steven Englehardt and Arvind Narayanan.
\newblock Online tracking: {A} 1-million-site measurement and analysis.
\newblock In {\em {ACM} {SIGSAC} {CCS'16}, Vienna, Austria, October 24--28,
  2016.}, pages 1388--1401. {ACM}, 2016.

\bibitem{guard}
Amin FaizKhademi, Mohammad Zulkernine, and Komminist Weldemariam.
\newblock {FPGuard}: Detection and prevention of browser fingerprinting.
\newblock In {\em {DBSec} 2015, Fairfax, VA, USA, July 13--15, 2015.}, volume
  9149 of {\em Lecture Notes in Computer Science}, pages 293--308. Springer,
  2015.

\bibitem{across}
Marjan Falahrastegar, Hamed Haddadi, Steve Uhlig, and Richard Mortier.
\newblock Tracking personal identifiers across the web.
\newblock In {\em {PAM} 2016, Heraklion, Greece, March 31 -- April 1, 2016.},
  volume 9631 of {\em Lecture Notes in Computer Science}, pages 30--41.
  Springer, 2016.

\bibitem{timing}
Edward~W. Felten and Michael~A. Schneider.
\newblock Timing attacks on web privacy.
\newblock In {\em {ACM} {CCS} 2000, Athens, Greece, November 1--4, 2000.},
  pages 25--32. {ACM}, 2000.

\bibitem{ws}
I.~Fette and A.~Melnikov.
\newblock {The WebSocket Protocol}.
\newblock {RFC} 6455, {RFC Editor}, December 2011.

\bibitem{http}
R.~Fielding, J.~Gettys, J.~Mogul, H.~Frystyk, L.~Masinter, P.~Leach, and
  T.~Berners-Lee.
\newblock {Hypertext Transfer Protocol -- {HTTP/1.1}}.
\newblock {RFC} 2616, {RFC Editor}, June 1999.

\bibitem{metric}
David Fifield and Serge Egelman.
\newblock Fingerprinting web users through font metrics.
\newblock In {\em {FC}'15, San Juan, Puerto Rico, January 26--30, 2015.},
  volume 8975 of {\em Lecture Notes in Computer Science}, pages 107--124.
  Springer, 2015.

\bibitem{fake}
Ugo Fiore, Aniello Castiglione, Alfredo~De Santis, and Francesco Palmieri.
\newblock Countering browser fingerprinting techniques: Constructing a fake
  profile with google chrome.
\newblock In {\em {NBiS} 2014, Salerno, Italy, September 10--12, 2014.}, pages
  355--360. {IEEE} Computer Society, 2014.

\bibitem{fingerprint}
Jason Franklin and Damon McCoy.
\newblock Passive data link layer 802.11 wireless device driver fingerprinting.
\newblock In {\em {USENIX} Security '06, Vancouver, BC, Canada, July 31 --
  August 4, 2006}. {USENIX} Association, 2006.

\bibitem{footprint}
Balachander Krishnamurthy and Craig~E. Wills.
\newblock Generating a privacy footprint on the internet.
\newblock In {\em {ACM} {SIGCOMM} {IMC} 2006, Rio de Janeriro, Brazil, October
  25--27, 2006.}, pages 65--70. {ACM}, 2006.

\bibitem{beast}
Pierre Laperdrix, Walter Rudametkin, and Benoit Baudry.
\newblock Beauty and the beast: Diverting modern web browsers to build unique
  browser fingerprints.
\newblock In {\em {IEEE} {S\&P} 2016, San Jose, CA, USA, May 22--26, 2016.},
  pages 878--894. {IEEE} Computer Society, 2016.

\bibitem{accurate}
Hoan Le, Federico Fallace, and Pere Barlet{-}Ros.
\newblock Towards accurate detection of obfuscated web tracking.
\newblock In {\em {IEEE} {MN} 2017, Naples, Italy, September 27--29, 2017},
  pages 1--6. {IEEE}, 2017.

\bibitem{raider}
Adam Lerner, Anna~Kornfeld Simpson, Tadayoshi Kohno, and Franziska Roesner.
\newblock Internet jones and the raiders of the lost trackers: An
  archaeological study of web tracking from 1996 to 2016.
\newblock In {\em {USENIX} Security '16, Austin, TX, USA, August 10--12, 2016.}
  {USENIX} Association, 2016.

\bibitem{hidden}
Timothy Libert.
\newblock Exposing the invisible web: An analysis of third-party http requests
  on 1 million websites.
\newblock {\em International Journal of Communication}, 9:18, October 2015.

\bibitem{3rd}
Jonathan~R. Mayer and John~C. Mitchell.
\newblock Third-party web tracking: Policy and technology.
\newblock In {\em {IEEE} {S\&P} 2012, San Francisco, California, {USA}, 21--23
  May 2012.}, pages 413--427, 2012.

\bibitem{catch}
Georg Merzdovnik, Markus Huber, Damjan Buhov, Nick Nikiforakis, Sebastian
  Neuner, Martin Schmiedecker, and Edgar~R. Weippl.
\newblock Block me if you can: {A} large-scale study of tracker-blocking tools.
\newblock In {\em {IEEE} {EuroS\&P} 2017, Paris, France, April 26--28, 2017},
  pages 319--333. {IEEE}, 2017.

\bibitem{horde}
Hassan Metwalley, Stefano Traverso, Marco Mellia, Stanislav Miskovic, and Mario
  Baldi.
\newblock The online tracking horde: {A} view from passive measurements.
\newblock In {\em {TMA} 2015, Barcelona, Spain, April 21--24, 2015.}, volume
  9053 of {\em Lecture Notes in Computer Science}, pages 111--125. Springer,
  2015.

\bibitem{javascript}
Keaton Mowery, Dillon Bogenreif, Scott Yilek, and Hovav Shacham.
\newblock Fingerprinting information in javascript implementations.
\newblock In {\em W2SP 2011, Oakland, CA, USA, May 26, 2011.}, volume~2, pages
  180--193, 2011.

\bibitem{canvas}
Keaton Mowery and Hovav Shacham.
\newblock Pixel perfect: Fingerprinting canvas in {HTML5}.
\newblock In {\em W2SP 2012, San Francisco, CA, USA, May 24, 2012.} IEEE
  Computer Society, May 2012.

\bibitem{engine}
Martin Mulazzani, Philipp Reschl, Markus Huber, Manuel Leithner, Sebastian
  Schrittwieser, Edgar Weippl, and FC~Wien.
\newblock Fast and reliable browser identification with javascript engine
  fingerprinting.
\newblock In {\em {W2SP} 2013, San Francisco, CA, USA, May 24, 2013.},
  volume~5, 2013.

\bibitem{lies}
Nick Nikiforakis, Wouter Joosen, and Benjamin Livshits.
\newblock Privaricator: Deceiving fingerprinters with little white lies.
\newblock In {\em {WWW} 2015, Florence, Italy, May 18--22, 2015}, pages
  820--830. {ACM}, 2015.

\bibitem{monster}
Nick Nikiforakis, Alexandros Kapravelos, Wouter Joosen, Christopher Kruegel,
  Frank Piessens, and Giovanni Vigna.
\newblock Cookieless monster: Exploring the ecosystem of web-based device
  fingerprinting.
\newblock In {\em {IEEE} {S\&P} 2013, Berkeley, CA, {USA}, May 19--22, 2013},
  pages 541--555. {IEEE} Computer Society, 2013.

\bibitem{battery}
Lukasz Olejnik, Gunes Acar, Claude Castelluccia, and Claudia D{\'{\i}}az.
\newblock The leaking battery --- {A} privacy analysis of the {HTML5} battery
  status {API}.
\newblock In {\em {DPM} {QASA} 2015, Vienna, Austria, September 21--22, 2015.},
  volume 9481, pages 254--263. Springer, 2015.

\bibitem{privacy}
Georgios Portokalidis, Michalis Polychronakis, Angelos~D. Keromytis, and
  Evangelos~P. Markatos.
\newblock Privacy-preserving social plugins.
\newblock In {\em USENIX Security '12, Bellevue, WA, {USA}, August 8--10,
  2012.}, pages 631--646. {USENIX} Association, 2012.

\bibitem{https}
Eric Rescorla.
\newblock {HTTP} over {TLS}.
\newblock {RFC} 2818, {RFC Editor}, May 2000.

\bibitem{defend}
Franziska Roesner, Tadayoshi Kohno, and David Wetherall.
\newblock Detecting and defending against third-party tracking on the web.
\newblock In {\em {USENIX} {NSDI} '12, San Jose, CA, USA, April 25--27, 2012.},
  pages 155--168. {USENIX} Association, 2012.

\bibitem{css}
Naoki Takei, Takamichi Saito, Ko~Takasu, and Tomotaka Yamada.
\newblock Web browser fingerprinting using only cascading style sheets.
\newblock In {\em {BWCCA} 2015, Krakow, Poland, November 4--6, 2015.}, pages
  57--63. {IEEE} Computer Society, 2015.

\bibitem{block}
Christof~Ferreira Torres, Hugo~L. Jonker, and Sjouke Mauw.
\newblock {FP-Block}: Usable web privacy by controlling browser fingerprinting.
\newblock In {\em {ESORICS} 2015, Vienna, Austria, September 21--25, 2015.},
  volume 9327 of {\em Lecture Notes in Computer Science}, pages 3--19.
  Springer, 2015.

\bibitem{shpf}
Thomas Unger, Martin Mulazzani, Dominik Fruhwirt, Markus Huber, Sebastian
  Schrittwieser, and Edgar~R. Weippl.
\newblock {SHPF:} enhancing {HTTP(S)} session security with browser
  fingerprinting.
\newblock In {\em {ARES} 2013, Regensburg, Germany, September 2--6, 2013.},
  pages 255--261. {IEEE} Computer Society, 2013.

\bibitem{classification}
Randika Upathilake, Yingkun Li, and Ashraf Matrawy.
\newblock A classification of web browser fingerprinting techniques.
\newblock In {\em {NTMS} 2015, Paris, France, July 27--29, 2015.}, pages 1--5.
  {IEEE}, 2015.

\bibitem{creepy}
Blase Ur, Pedro~Giovanni Leon, Lorrie~Faith Cranor, Richard Shay, and Yang
  Wang.
\newblock Smart, useful, scary, creepy: perceptions of online behavioral
  advertising.
\newblock In {\em {SOUPS} '12, Washington, DC, {USA}, July 11--13, 2012.},
  page~4. {ACM}, 2012.

\bibitem{stalker}
Antoine Vastel, Pierre Laperdrix, Walter Rudametkin, and Romain Rouvoy.
\newblock {FP-STALKER}: Tracking browser fingerprint evolutions.
\newblock In {\em {IEEE} {S\&P} 2018, San Fransisco, CA, USA, May 21--23,
  2018}, pages 1--14. IEEE, 2018.

\bibitem{private}
Bin Zhao and Peng Liu.
\newblock Private browsing mode not really that private: Dealing with privacy
  breach caused by browser extensions.
\newblock In {\em {IEEE/IFIP} {DSN} 2015, Rio de Janeiro, Brazil, June 22--25,
  2015.}, pages 184--195, 2015.

\bibitem{analysis}
Sebastian Zimmeck, Jie~S. Li, Hyungtae Kim, Steven~M. Bellovin, and Tony
  Jebara.
\newblock A privacy analysis of cross-device tracking.
\newblock In {\em {USENIX} Security '17, Vancouver, BC, Canada, August 16--18,
  2017.}, pages 1391--1408. {USENIX} Association, 2017.

\end{thebibliography}
	\appendix
	\section*{Appendices}
	\section{Crawling Components and Environment}
	\subsection{Prepopulated List of Attributes} \label{prepopulate}
	Resolution, OS, OS Version, User-Agent, Browser Name, Browser Version, WebGL Renderer, WebGL Vendor, WebGL Version, GPU, GPU Vendor, Installed Plugins, Language, Geolocation, City, IP Addresses, and Charset.
	\subsection{ Crawler Software Components }
	
	\begin{center}
		\begin{table}[H]
			\label{crawl}
			\begin{center}
				\begin{tabular}{|l|l|}
					\hline
					\rowcolor[HTML]{ EFEFEF}
					\textbf{Component} & \textbf{Details}
					\\ \hline
					Browser extension & FingerprintAlert 1.0
					\\ \hline
					
					Programming language & Phython 3.6.3
					
					\\ \hline
					Automation tool & Selenium 3.8.1
					
					\\ \hline
					
				\end{tabular}
				
			\end{center}		
		\end{table}
	\end{center}

\subsection{Computing Environment} \label{config}

	\begin{center}
		\begin{table}[H]
			\begin{center}
				\begin{tabular}{|l|l|}
					\hline
					\rowcolor[HTML]{ EFEFEF}
					\textbf{Component} & \textbf{Details}
					\\ \hline
					Device & MacBook Pro (10.1.1)
					\\ \hline
					
					OS & MacOS Sierra 12.1
					
					\\ \hline
					Browser & Chrome 62.0.3202.94
					
					\\ \hline
					
				\end{tabular}
				
			\end{center}		
		\end{table}
	\end{center}
\section{Attributes Collected by Fingerprinters} \label{full}

\subsection{WebGL} \label{append:webgl}
\texttt{\seqsplit{aliased line width range, aliased point size range, alpha bits, angle instanced arrays, antialiasing, blue bits, depth bits, experimental-webgl, ext blend min max, ext disjoint timer query, ext frag depth, ext shader texture lod, ext srgb, ext texture filter anisotropic, fragment shader high float precision, fragment shader high float precision range max, fragment shader high float precision range min, fragment shader high int precision, fragment shader high int precision range max, fragment shader high int precision range min, fragment shader low float precision, fragment shader low float precision range max, fragment shader low float precision range min, fragment shader low int precision, fragment shader low int precision range max, fragment shader low int precision range min, fragment shader medium float precision, fragment shader medium float precision range max, fragment shader medium float precision range min, fragment shader medium int precision, fragment shader medium int precision range max, fragment shader medium int precision range min, green bits, max 3d texture size, max anisotropy, max array texture layers, max color attachments, max combined fragment uniform components, max combined texture image units, max combined vertex uniform components, max cube map texture size, max draw buffers, max fragment input components, max fragment uniform blocks, max fragment uniform components, max fragment uniform vectors, max program texel offset, max render buffer size, max samples, max texture image units, max texture lodbias, max texture size, max transform feedback interleaved components, max transform feedback separate attribs, max transform feedback separate components, max uniform block size, max uniform buffer bindings, max varying components, max varying vectors, max vertex attribs, max vertex output components, max vertex texture image units, max vertex uniform blocks, max vertex uniform components, max vertex uniform vectors, max view port dims, min program texel offset, oes element index uint, oes standard derivatives, oes texture float, oes texture float linear, oes texture half float, oes texture half float linear, oes vertex array object, performance caveat, red bits, renderer, shading language version, stencil bits, unmasked renderer webgl, unmasked vendor webgl, vendor, version, vertex shader high float precision, vertex shader high float precision range max, vertex shader high float precision range min, vertex shader high int precision, vertex shader high int precision range max, vertex shader high int precision range min, vertex shader low float precision, vertex shader low float precision range max, vertex shader low float precision range min, vertex shader low int precision, vertex shader low int precision range max, vertex shader low int precision range min, vertex shader medium float precision, vertex shader medium float precision range max, vertex shader medium float precision range min, vertex shader medium int precision, vertex shader medium int precision range max, vertex shader medium int precision range min, webgl, webgl compressed texture s3tc, webgl compressed texture s3tc srgb, webgl debug renderer info, webgl debug shaders, webgl depth texture, webgl draw buffers, webgl lose context, webgl2, webkit ext texture filter anisotropic, webkit webgl compressed texture s3tc, webkit webgl depth texture, webkit webgl lose context.}}
\subsection{Features}
\texttt{\seqsplit{adblock, application cache, background size, blending, bluetooth, border image, border radius, box shadow, canvas, canvas webp, canvas blending, canvas winding, credentials, css animations, css columns, css gradients, css reflections, css transforms, css transforms 3dc, css transitions, drag and drop, flex box, flex box legacy, font face, generated content, get battery, get game pads, get user media, hash change, history, hsla, img hash, inline svg, installed fonts, installed plugins, java enabled, js, media decvices, mime types, multiple bgs, opacity, permissions, post message, presentation, register protocol handler, request media key system access, request midi access, rgba, send beacon, service worker, shockwave flash, smil, svg, svg clip paths, text shadow, towebp, unregister protocol handler, usb, vibrate, web sql database, web workers, webkit get user media, webkit persistent storage, webkit temporary storage, webrtc, websockets.}}

\subsection{Media}
\texttt{\seqsplit{ac-base latency, ac-channel count, ac-channel count mode, ac-channel interpretation, ac-max channel count, ac-number of inputs, ac-number of outputs, ac-sampler ate, ac-state, an-channel count, an-channel count mode, an-channel interpretation, an-fft size, an-frequency bin count, an-max decibels, an-min decibels, an-number of inputs, an-number of outputs, an-smoothing time constant, audio ogg, avc1.42c00d, avc1.42e01e (mp4a.40.2), codecs1, dynamiccompressor, h264, hybridoscillator, mp3, mp4a.40.2, mpeg, opus, oscillator, theora, video mp4, video ogg, vorbis (ogg), vorbis (vp8), vorbis (vp9), vorbis (wav), wav, webm, wm4a.}}

\subsection{Input/Output}
\texttt{\seqsplit{windowstate, outerheight, outerwidth, innerheight, innerwidth, width, height, availablewidth, availableheight, colordepth, keytimes, mouse, orientation, scrolls, maxtouchpoints, touchevent, touchstart, speakersinstalled, webcamsinstalled, microphonesinstalled.}}
\subsection{Network}
\texttt{\seqsplit{downlink, effectivetype, is proxied, is tor, is using tor exit node, local ip, onchange, public ipv4, public ipv6, rtt.}}	

\subsection{Miscellaneous}
\texttt{\seqsplit{app code name, battery level, charging, charging time, charset, collect time, cookie enabled, cpu cores, discharging time, do not track, geolocation, graphics card vendor, hardware concurrency, has timezone mismatch, incognito, indexed db, js heap size limit, languages, local storage, navigator, online, open data base, platform, product, product sub, referrer, renderer, session storage, timestamp, timezone, total js heap size, used js heap size, user agent, vendor, vendor sub.}}

\end{document}